\documentclass[article,twocolumn,amsmath,floats,showpacs,nofootinbib]{revtex4}
\usepackage{graphicx}
\usepackage{textcomp}

\usepackage{hyperref}

\hypersetup{colorlinks=true}

\begin{document}

\title{Inelastic electron-phonon scattering and excess current in superconducting point
contacts with a short coherence length}
\author{ N. L. Bobrov}
\affiliation{B.I.~Verkin Institute for Low Temperature Physics and
Engineering, of the National Academy of Sciences
of Ukraine, prospekt Lenina, 47, Kharkov 61103, Ukraine
Email address: bobrov@ilt.kharkov.ua}
\published {(\href{http://fntr.ilt.kharkov.ua/fnt/pdf/41/41-8/f41-0768r.pdf}{Fiz. Nizk. Temp.}, \textbf{41}, 768 (2015)); ( \href{http://dx.doi.org/10.1063/1.4929594}{Low Temp. Phys.}, \textbf{41}, 595 (2015)}
\date{\today}

\begin{abstract}Nonlinear electrical effects in superconducting $S-c-S$ contacts, including the spectroscopy of
electron-phonon interactions (EPI) in these systems, and the recovery of the EPI function from
experimental data are discussed. The effect of a magnetic field on the current-voltage characteristics
(I-V curves) and their derivatives for $ErNi_{2}B_{2}C$ point contacts (PC) with $d\ge \xi $ (where $d$ is the
diameter of the PC and $\xi $ is the coherence length) is studied. It is found that in zero magnetic fields
and in near-critical fields, when the size of the superconducting gap can be neglected, the position
of the peaks in $dV/dI$ coincides with the peaks in the Yanson EPI spectra. In low fields the peaks
are shifted toward lower energies and in intermediate fields, the peaks split. For PC with diameters
greater than or on the order of the coherence length, the relative size of the negative phonon contribution
to the excess current is considerably greater than in ballistic contacts. This leads to substantial
suppression of the high-frequency peaks in the spectra for the superconducting state. In order to
recover the EPI function from these spectra it is necessary to correct their intensities at high energies.
For "dirty" $NbSe_2$ and $Nb$ point-contacts with $d\ge \xi $, which have no phonon features in the
second derivative of the I-V curve in the normal state, the EPI can be reconstructed from the superconducting
state.
\end{abstract}
\maketitle
\section{INTRODUCTION}
\subsection{Basic theoretical concepts}
In ballistic point contacts, an electron passing through the
short joining the metallic electrodes acquires an energy $eV$ from
the applied voltage. At any point in its trajectory it can lose
excess energy and emit a nonequilibrium phonon. If this occurs
in the immediate vicinity of the short, then there is a fairly high
probability that the electron will return to the electrode from
which it emerged. In this case, the resistance of the pointcontact
(PC) is supplemented by an additional nonlinear contribution.
Although only a small fraction of the electrons are scattered
on nonequilibrium phonons in the neighborhood of the
short and only some of these scattered electrons turn back, the
deviation from Ohm's law owing to the high current density
($\sim 1\cdot {{10}^{8}}\ \text{A/c}{{\text{m}}^{\text{2}}}$) in the constriction can be very substantial
and can approach ten percent. The main task of Yanson PC
spectroscopy is the study of this nonlinearity, since the second
derivative of the I-V curve of a PC in the normal state is proportional
to the electron-phonon interaction (EPI) function ${{G}_{pc}}(\omega )$ \cite{Kulik}:
\begin{equation}
\label{eq__1}
{{G}_{pc}}\left( \omega  \right)={{\left. -\frac{3{{R}_{0}}\hbar {{v}_{F}}}{32ed}\cdot \frac{{{d}^{2}}I}{d{{V}^{2}}} \right|}_{\omega ={eV}/{\hbar }\;}}
\end{equation}
The distance over which most of these electrons lose
excess energy is referred to as the energy relaxation length
and depends strongly on the applied voltage. The minimum
energy relaxation length is attained when the energy of an
electron is comparable to or exceeds the maximum possible
energy of a phonon at the edge of the spectrum.
If one or both of the electrodes forming a point-contact
is superconducting, then an additional transport channel
owing to the presence of excess current shows up in the
PC conductivity through the constriction. When the voltage
on the contact is considerably higher than the superconducting
gap, and when all the nonlinearities in the $I(V)$
curve are taken into account, the current can be written in
the form \cite{Khlus}
\begin{equation}
\label{eq__2}
I(V)=\frac{V}{R}+\delta I_{ph}^{N}(V)+I_{exc}^{0}+\delta I_{ph}^{S}(V).
\end{equation}
Here $\delta I_{ph}^{N}(V)$ is the nonlinearity owing to the EPI in a normal
PC and is the basis of Yanson spectroscopy (see Eq. (\ref{eq__1})).
When $eV>>\Delta $, the excess current for a ballistic $S-c-S$ PC is
independent of the bias and is given by \cite{Zaitsev}:
\begin{equation}
\label{eq__3}
I_{exc}^{0}=\frac{8\Delta }{3eR}th\frac{eV}{2T},
\end{equation}
and, finally, $\delta I_{ph}^{S}(V)$ is the negative contribution to the excess
current owing to collisions of nonequilibrium phonons with
Andreev electrons (i.e., quasi-electrons, during the process
of electron-hole conversion). Because of these collisions, the
number of Andreev electrons is reduced, and this leads to a
drop in the excess current. For the excess current of an $S-c-S$
contact we have \cite{Khlus}
\begin{equation}
\label{eq__4}
\frac{d{{I}_{exc}}}{dV}=-\frac{64}{3R}\left( \frac{\Delta L}{\hbar \bar{v}} \right){{\left[ {{G}^{N}}(\omega )+\frac{1}{4}{{G}^{S}}(\omega ) \right]}_{\omega =eV/\hbar }}
\end{equation}
Here $\bar{v}$ is the velocity of the electrons averaged over the
Fermi surface. On comparing this with Eq. (\ref{eq__1}), it is clear that
here the EPI function is proportional to the first derivative.
The relative magnitude of the negative phonon contribution
to the excess current for ballistic contacts near the
Debye energies is on the order of \cite{Khlus}
\begin{equation}
\label{eq__5}
\delta I_{ph}^{S}(V)\sim {d\cdot {{\omega }_{D}}}/{{{v}_{F}}}\;
\end{equation}
i.e., it is small compared to the bias-independent excess
current.

The relative smallness of this contribution is caused by
the ballistic character of the PC, since the probability of
inelastic scattering of nonequilibrium phonons on Andreev
electrons depends on their mutual concentration. The maximum
concentration coincides with the region of maximum
current density and falls off rapidly with distances from the
constriction. Thus, in ballistic contacts the volume for generation
of the phonons that form the PC spectrum in the normal
state is close to the volume within which the spectrum
owing to the suppression of the excess current is formed.
Because of this, the EPI functions recovered from the characteristics
of these contacts in the normal state are extremely
close to the EPI functions obtained from the excess current
(see Figs. 2, 5 and 6 in Ref. \cite{Bobrov1}). There the inelastic superconducting
contribution to the spectrum in the region of the phonon
energies for $S-c-N$ point-contacts shows up in the form
of \emph{differential resistance peaks} in the first derivative of the
excess current, \emph{shifted} toward lower energies by an amount
on the order of the gap, while for $S-c-S$ contacts there is no
shift, i.e., these differences are minimal at low temperatures.
On the other hand, if the condition that the diameter of
the PC be small compared to the coherence length and to the
energy relaxation length of the electrons at the Debye energies
is not satisfied rigorously enough, then the volume
within which the spectrum in the superconducting state is
formed may increase because part of the region near the contact
is drawn in owing to an increase in the concentration of
Andreev electrons and nonequilibrium phonons there. In this
case, the spectra obtained in the normal state and recovered
from the excess current may differ. This is the case of greatest
practical interest. When PC are produced by mechanical
methods, the largest distortions of the crystal lattice are concentrated
in the surface layer at the point of contact of the
electrodes. In addition, the surface itself is generally of inferior
quality compared to the volume. Thus, it is extremely
difficult to obtain a ballistic point-contact. Sometimes the
spectrum in the normal state does not have any phonon features
because of amorphization of the material in the contact
region. If the distortions affect the volume only in the center
of the contact and do not extend into the depth of the material
being studied, it may be possible to recover the EPI successfully
from the negative contribution to the excess
current \cite{Bobrov2}. In this case, it is obvious that the volumes responsible
for formation of the spectra in the normal and superconducting
states do not coincide in space. The phonon structure
in the superconducting state is produced by a larger region
near the contact with a more perfect crystal lattice.
For many superconductors, distortions in the lattice
cause reductions in several of the superconducting parameters
or even complete suppression of superconductivity.
Then it becomes possible to obtain EPI spectra from the
most perfect superconducting regions in the cantilever near
the short.

If the condition that the contact diameter should be small
compared to the coherence length is not strict, i.e., $d\le {{\xi }_{0}}$,
then the EPI spectrum recovered from the excess current for
a PC with an amorphous core will have a shape close to the
spectrum for a ballistic contact in the normal state.

Condition (\ref{eq__5}) means that the number of Andreev electrons
varies little up to the Debye energies; this ensures a
proper form for the EPI spectrum recovered from the excess
current. When the excess current is strongly suppressed by
nonequilibrium phonons, the shape of the spectrum can be
highly distorted. The phonon peaks at low biases will be
emphasized, since the excess current is not strongly suppressed
and the concentration of Andreev electrons is high.
And so high-energy phonons can appear much weaker when
the excess current is reduced substantially. This may require
correction of the high-frequency part of the spectrum with the
energy dependence of the excess current taken into account.
This applies, first of all, to superconductors with a low
electron energy relaxation length at the Debye energies, as
well as with a short coherence length. In contacts between
these superconductors, inelastic scattering of nonequilibrium
phonons on Andreev electrons in the cantilevers near the
short may become an important source of information on the EPI.

\section{Recovery of EPI functions}
\subsection{$\text{ErN}{{\text{i}}_{\text{2}}}{{\text{B}}_{\text{2}}}\text{C}$ point-contacts}

\begin{figure}[tbp]
\includegraphics[width=8.5cm,angle=0]{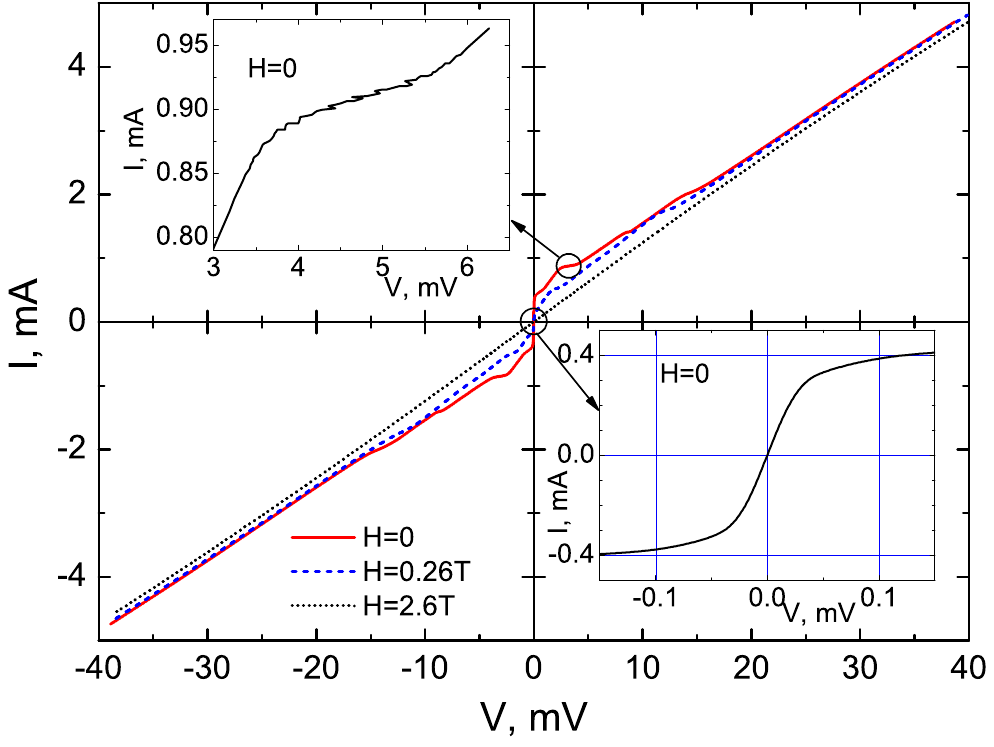}
\caption[]{I-V curves of an $\text{ErN}{{\text{i}}_{\text{2}}}{{\text{B}}_{\text{2}}}\text{C}$ point-contact in different magnetic
fields: $R_{0}^{N}=8\ \Omega$, $\text{T}\approx \text{1}.\text{6}\ ~\text{K}$. The inset at the upper left shows the instability
region of the curve. The inset at the lower right shows the initial section of
the curve in zero magnetic field, $R_{0}^{S}=0.09\ \Omega$.}
\label{Fig1}
\end{figure}

\begin{figure}[]
\includegraphics[width=8.5cm]{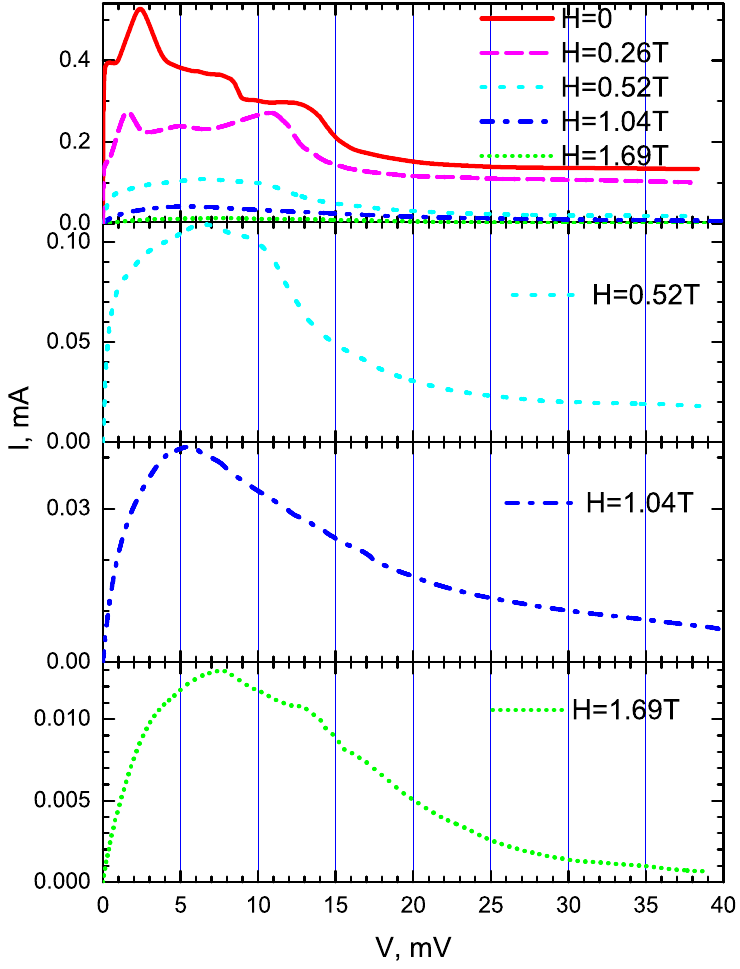}
\caption{The excess current of the point-contact shown in Fig. \ref{Fig1} for different
magnetic fields. For all fields the relative magnitude of the negative phonon
contribution to the excess current is large compared to the bias-independent
excess current.}
\label{Fig2}
\end{figure}

The influence of antiferromagnetic ordering on the
superconducting gap in the nickel borocarbide superconductor
$ErNi_2B_2C$ has been examined in detail elsewhere \cite{Bobrov3,Bobrov4} and
the experimental technique is described in detail there. In
this section, we examine the effect of nonequilibrium phonons
on the excess current.
Measurements were made at $T\approx 1.6\ {K}$. Figure \ref{Fig1}
shows the I-V curves for an $S-c-S$ contact in different magnetic
fields. The resistance of the PC is $R_{0}^{N}=8\ \Omega $ and the diameter
estimated using the Veksler formula (see p. 9 of Ref. \cite{Naidyuk})
\begin{equation}
\label{eq__6}
d\simeq \frac{\rho }{2R}+\sqrt{{{\left( \frac{\rho }{2R} \right)}^{2}}+\frac{16\rho l}{3\pi R}}
\end{equation}
is $d$=17 nm, where it is assumed that $\rho l\cong {{10}^{11}}\ \Omega \cdot c{{m}^{2}}$
(Ref. \cite{Shulga}) and $\rho \sim \text{3}.\text{5 }\cdot \text{1}{{0}^{-\text{6}}}\Omega \ \text{cm}$
\cite{Cho}. The elastic mean free
path is then ${{l}_{i}}\sim{\ }\text{28}.\text{5}~\ \text{nm}$ and the coherence length is
${{\xi }_{0}}=\text{15}~\text{nm}$ \cite{Skanthakumar}. Thus, the diameter of the contact is less than
the coherence length. Since the contact is connected in a
quasi-four contact circuit, the contact resistance in zero magnetic
field is $R_{0}^{S}\simeq 0.09\ \Omega $ (see the inset to Fig. \ref{Fig1}). The dependence
of the resistance for zero bias is shown in Fig. \ref{Fig2}
and indicates that the PC is stable in a magnetic field. For
biases of 3.5-6 $mV$, the I-V curve is found to be unstable
(see the inset to Fig. \ref{Fig1}), but there is no hysteresis. Similar
behavior has been observed in an $NbS{{e}_{2}}-Cu$ PC \cite{Yanson1}.  It is
impossible to record the derivative in this interval; the recorder
pen produces a chaotic trace. In order to match the derivative
before and after the stability, we found a polynomial
fit through this segment of the I-V curve and obtained the derivative
by numerical integration. This instability appears to
be related to the smallness of ${{\xi }_{0}}$ relative to the PC diameter.
Before a certain density of the transport current is reached,
the superconducting boundary moves from the constriction
into the depth of the cantilevers \cite{Yanson1}. Note that for $T\le 2\ K$ the
critical magnetic field ${{H}_{{{c}_{2}}}}\le 2.4\ T$ \cite{Bud'ko}, which is somewhat
lower than the maximum field attained here, but, as will be
shown below, residual traces of superconductivity still show
up. Nevertheless, for finding the excess current, the I-V
curve for a field of $H=2.6\ T$ was used as the normal state.
\begin{figure}[]
\includegraphics[width=8.5cm,angle=0]{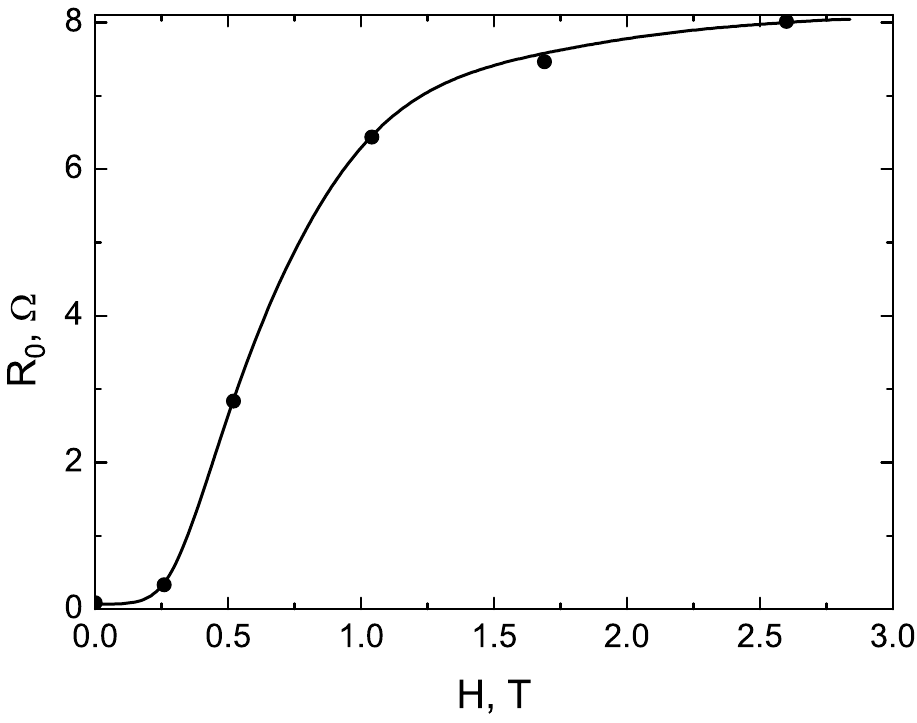}
\caption[]{Resistance of an $\text{ErN}{{\text{i}}_{\text{2}}}{{\text{B}}_{\text{2}}}\text{C}$
PC as a function of magnetic field for zero bias.}
\label{Fig3}
\end{figure}
Figure \ref{Fig3} shows the excess current as a function of bias on a
point-contact in different magnetic fields. Taking the average
effective gap to be $\Delta=1.5\ mV$ (Ref. \cite{Bobrov3}) yields estimates for
the "clean" limit of ${{I}_{exc}}=0.5\ mA$ and for the "dirty" limit,
${{I}_{exc}}=0.275\ mA$. The experimental value according to the
data of Fig. \ref{Fig3} is ${{I}_{exc}}=0.36\ mA$, i.e., it lies roughly midway
between the "clean" and "dirty" limits. The critical current
${{I}_{c}}=0.39\ mA$ coincides precisely with the "dirty" limit. (In
the clean limit ${{I}_{c}}=0.59\ mA$.) Since $I_c$ depends strongly on
the magnetic field, the lack of shielding of the point-contact
causes this difference. Thus, the contact is in an intermediate
electron drift regime.

\begin{figure}[]
\includegraphics[width=8.5cm,angle=0]{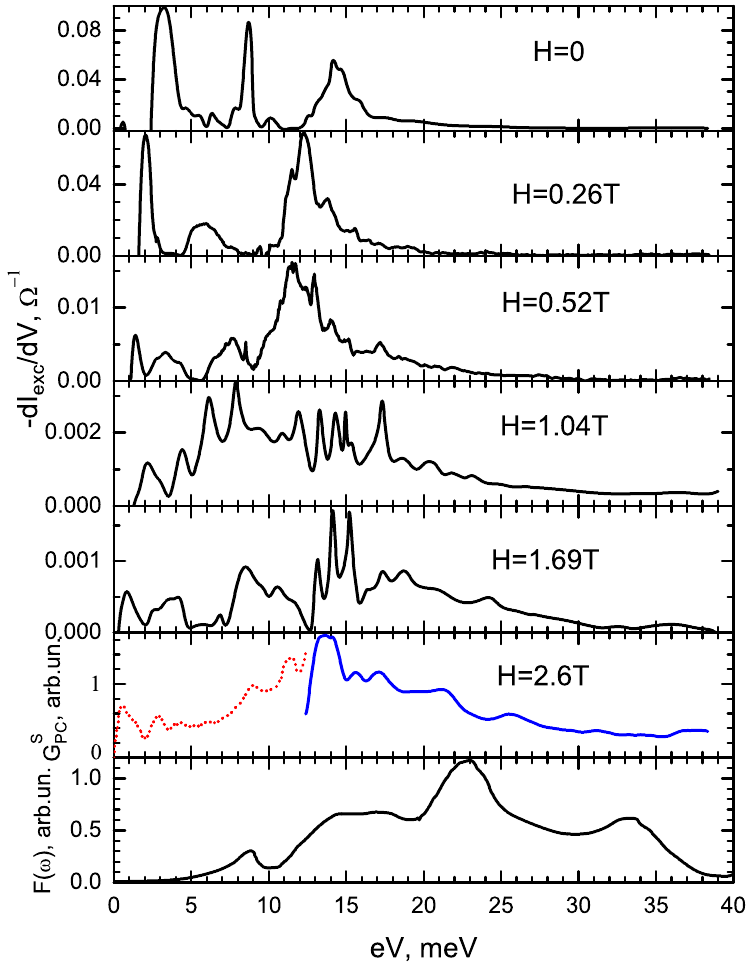}
\caption[]{The first derivatives of the excess current in an $ErN{{i}_{2}}{{B}_{2}}C$ pointcontact,
which are proportional to the EPI function (Eq. (\ref{eq__4})), in different
magnetic fields, and the phonon state density function. $F\left( \omega  \right)$ and $G_{pc}^{S}$
measured in arbitrary units.}
\label{Fig4}
\end{figure}

Figure \ref{Fig2} shows that, even for energies below the Debye
energies, the relative magnitude of the negative phonon contribution
to the excess current is extremely large, i.e., condition
(\ref{eq__5}) does not hold. This is related to the failure of the
condition $d\ll {{\xi }_{0}}$. Since the strong drop in the excess current
even at the first two peaks in the density of phonon states
leads to a significant reduction in the concentration of
Andreev electrons, this is accompanied by a corresponding
reduction in the intensity of the subsequent phonon modes in
the spectra. Figure \ref{Fig4}
shows the first derivatives of the excess
current in different magnetic fields, which are proportional
to the EPI function (see Eq. (\ref{eq__4})), as well as the density function
of the phonon states. In zero magnetic field only two
peaks near 9 and $14\ mV$ show up; no higher frequency peaks
can be seen against the background of these. For high biases,
the traces were taken with the same modulation, so it was
not possible to recover the high-frequency part, as is done in
the next section, which is devoted to $NbSe_2$.

Recall that for $S-c-N$ point-contacts the peaks in the first
derivative of the excess current are shifted relative to the
EPI function by an amount somewhat less than $\Delta$ (see Figs.
1 and 2 of Ref. \cite{Bobrov1}), while for $S-c-S$ contacts there is no shift
(Figs. 5 and 6 of Ref. \cite{Bobrov1}).

In a magnetic field, the type-II superconductor is penetrated
by vortices, i.e., normal (the core of a vortex) and
superconducting regions coexist in the neighborhood of the
contact. This problem has not been examined theoretically,
and experimental data are shown here for the first time.
Figure \ref{Fig4} shows that in fields $H=0.26$ and $0.52\ T$ there
is a significant shift in the peaks toward lower energies; the
shift is considerably larger than in the spectra of ballistic
$S-c-N$ point-contacts in zero field.

Further increases in the field to $H=1.04$ and $1.69\ T$
cause splitting of the peaks, which begin to gravitate toward
the position characteristic of zero field. The nature of this
behavior of the spectrum is not understood, but it does suggest
that it would be extremely difficult to recover the EPI
function for $S-c-S$ point-contacts in a magnetic field. We
have observed the strong effect of a magnetic field on the
EPI spectra of Ta ballistic point-contacts in the superconducting
state before \cite{Yanson2}.

\begin{figure}[]
\includegraphics[width=8.5cm,angle=0]{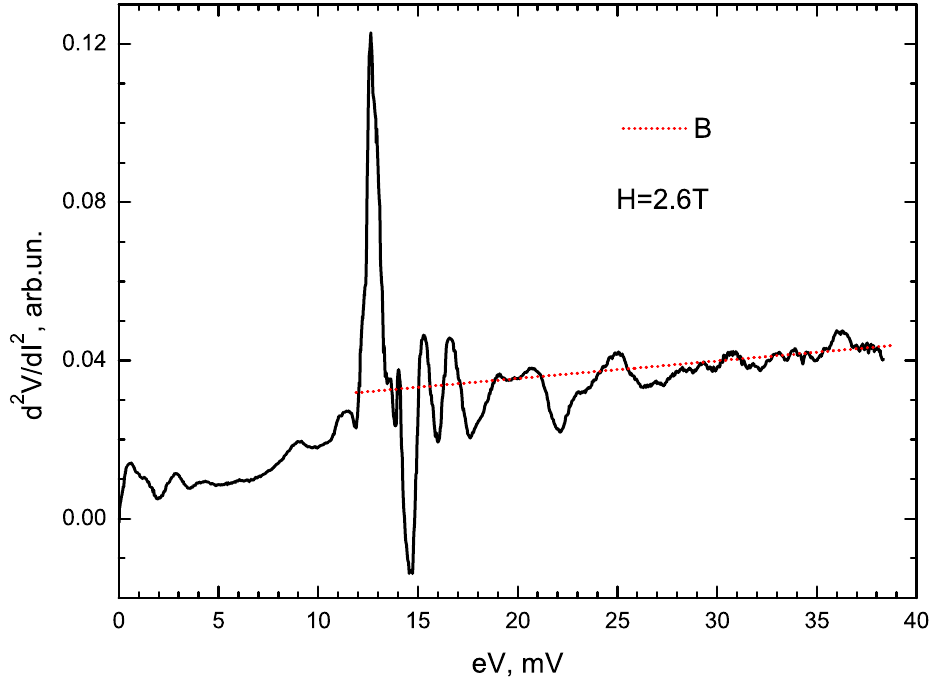}
\caption[]{The second derivative of the I-V curve of the $ErN{{i}_{2}}{{B}_{2}}C$ pointcontact
(shown in Fig. \ref{Fig1}) in a magnetic field. The dotted line shows the
assumed background.}
\label{Fig5}
\end{figure}

We now examine the spectrum in a field of $H=2.6\ T$,
near the complete suppression of superconductivity, in more
detail. The initial segment (the dotted line) up to about
$13\ meV$ is the second derivative of the I-V curve without
changes. The next part is recovered from the second
derivative of the I-V curve after subtracting the background
and integrating (Fig. \ref{Fig5}).
Since the probability of scattering of
nonequilibrium phonons on Andreev electrons depends on
their concentration, and the phonon concentration increases
especially rapidly near the peaks, the critical concentration
of these needed for the appearance of peaks in the derivative
of the I-V curve is reached for biases starting at the second
phonon peak.

\subsection{Point-contacts based on $NbSe_2$}
We have published data on EPI in ${NbS}{{{e}}_{{2}}}$ previously \cite{Bobrov2}.
Here we consider a point-contact with a large diameter in
which the condition ${d}\ll {{\xi }_{0}}$  is not met. Although the contact
has a rather high resistance in terms of point-contact spectroscopy
$R_{0}^{N}=55\ \Omega $, its diameter is large at $d\approx 17\text{ }nm$,
given the electronic parameters $\rho l=2.2\times {{10}^{-11}}\ \Omega \cdot c{{m}^{2}}$ and
${{\rho }_{res}}=6.7\times {{10}^{-6}}\ \Omega \cdot cm$. Here the coherence length is ${{\xi }_{0}}=7.7\ nm$, or less than half the contact diameter. As a comparison,
the diameter of the $R_{0}^{N}=1000\ \Omega$ point-contact
examined in Ref. \cite{Bobrov2} is $d\approx 2\ nm$. Although the PC is of an $S-c-S$ type, there is no critical current and the excess current is about $8\ \mu A$, which is substantially below the "dirty" limit.
\begin{figure}[]
\includegraphics[width=8.5cm,angle=0]{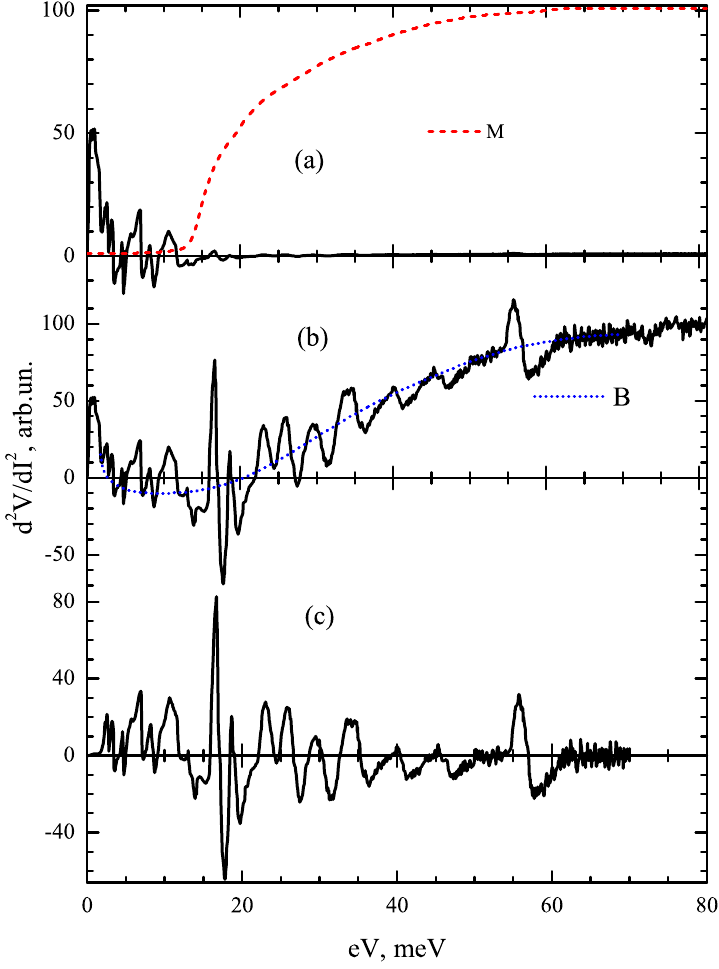}
\caption[]{\\(a) Spectrum of an $NbS{{e}_{2}}-NbS{{e}_{2}}$ point-contact (see Fig. 5 of Ref.
\cite{Bobrov5}). $R_{0}^{N}=54\ \Omega $ and $M$ is the multiplicative scaling curve. The scaling curve
was obtained from part of the plot of the excess current as a function of
energy (Fig. \ref{Fig7}). This part was turned by ${{180}^{\circ }}$, shifted toward higher energies
by 3~mV, and normalized so that ${{M}_{\min }}=1$ and ${{M}_{\max }}=100$. \\(b) The spectrum
after multiplication by the scaling factor, $B$ is the background curve.\\(c) The spectrum after subtraction of the background curve.}
\label{Fig6}
\end{figure}
The spectrum shown in Fig. \ref{Fig6}(a) is distinctly different
from that of the high-resistance PC shown in Fig. \ref{Fig3} of Ref. \cite{Bobrov2}.
Here the spectrum falls off very rapidly in intensity and by
$20\ mV$ it degenerates into a line. At the same time, the spectrum
of the high-resistance contact has an almost constant intensity
up to the Debye energy $\sim 60\ mV$. The damping of the
intensity of the spectrum correlates clearly with the damping
of the excess current in the PC (Fig. \ref{Fig7});
\begin{figure}[t]
\includegraphics[width=8.5cm,angle=0]{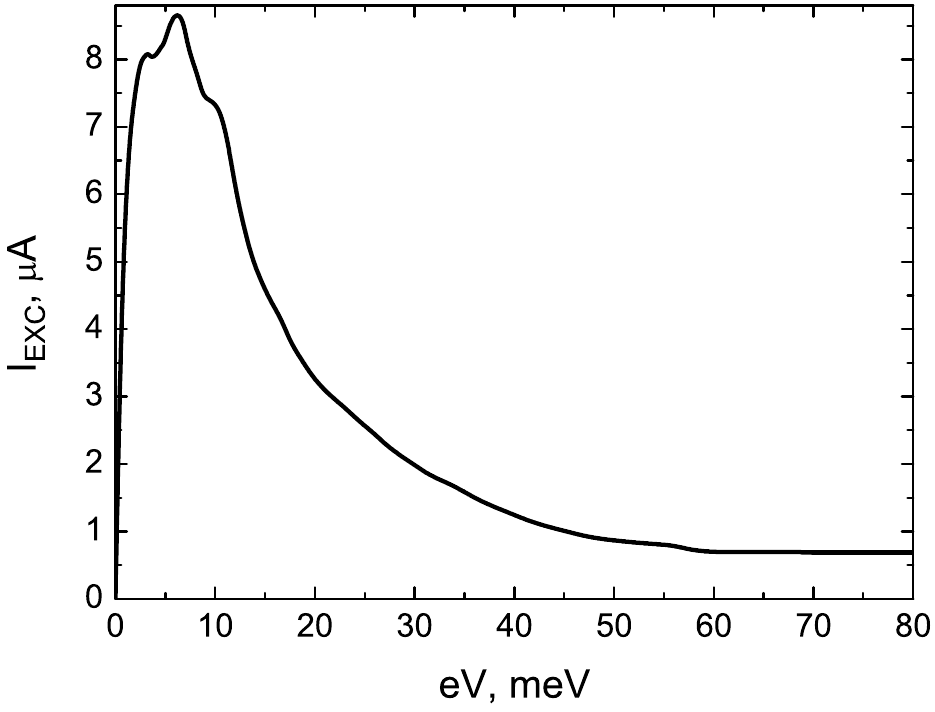}
\caption[]{Excess current of an $NbS{{e}_{2}}-NbS{{e}_{2}}$ point-contact as a function of
bias.}
\label{Fig7}
\end{figure}
thus, the scaling curve
$M$, which is based on the excess current curve in the
$\sim 10-80\ mV$ range, was used to correct this damping. The initial
segment of $M$ was chosen to be close to a straight line.
The excess current curve was turned by ${{180}^{\circ }}$ and shifted toward
higher energies by $3\ mV$. This curve was normalized so
that ${{M}_{\min }}=1$ and ${{M}_{\max }}=100$. Figure \ref{Fig6}(b) shows the corrected
spectrum after multiplication by the scaling curve $M$,
as well as the background curve $B$. Finally, Fig. \ref{Fig6}(c) is the
spectrum after subtraction of the background curve.
\begin{figure}[]
\includegraphics[width=8.5cm,angle=0]{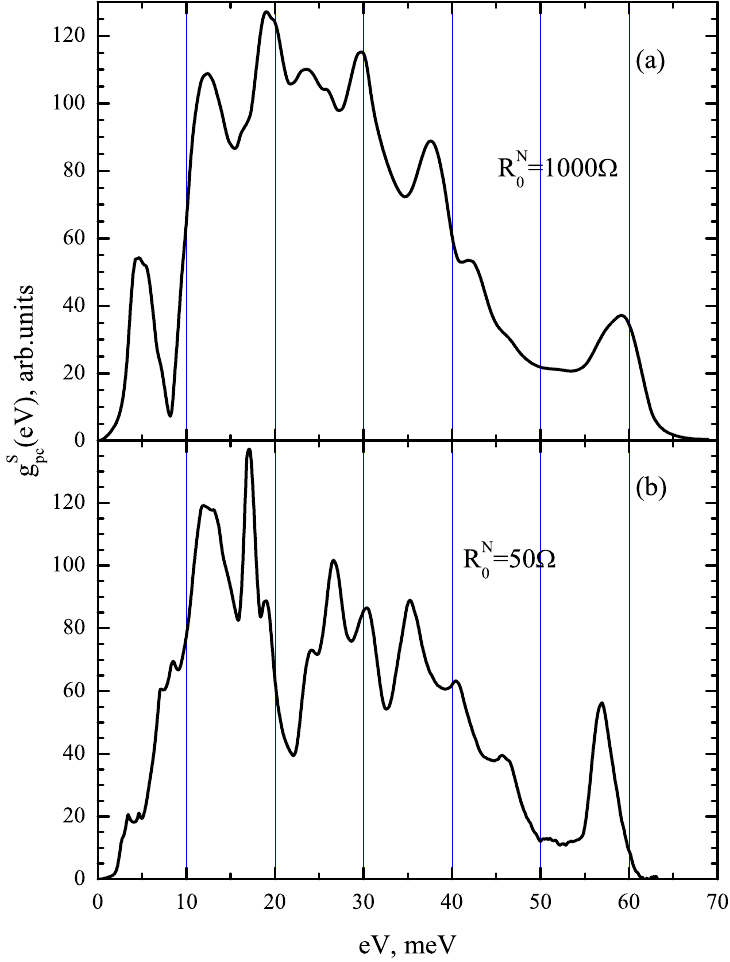}
\caption[]{\\(a) The EPI function recovered from a superconducting spectrum of
$NbS{{e}_{2}}-Cu$ (see Fig. 3 of Ref. \cite{Bobrov2}); \\(b) the EPI function recovered by integrating
the curve of Fig. \ref{Fig6}(c).}
\label{Fig8}
\end{figure}
Figure \ref{Fig8} shows the EPI functions recovered from the
high-resistance and low-resistance point-contacts. Thus, after correction their shapes are in good agreement, especially
given the high lability of point-contact spectra \cite{Bobrov5}. Note
that suppression of the excess current is not connected in any
way with ohmic heating, as indicated by the absence of
smearing of the EPI function at high frequencies compared
to the high-resistance contact.
\subsection{Point-contact based on Nb}
\begin{figure}[t]
\includegraphics[width=8.5cm,angle=0]{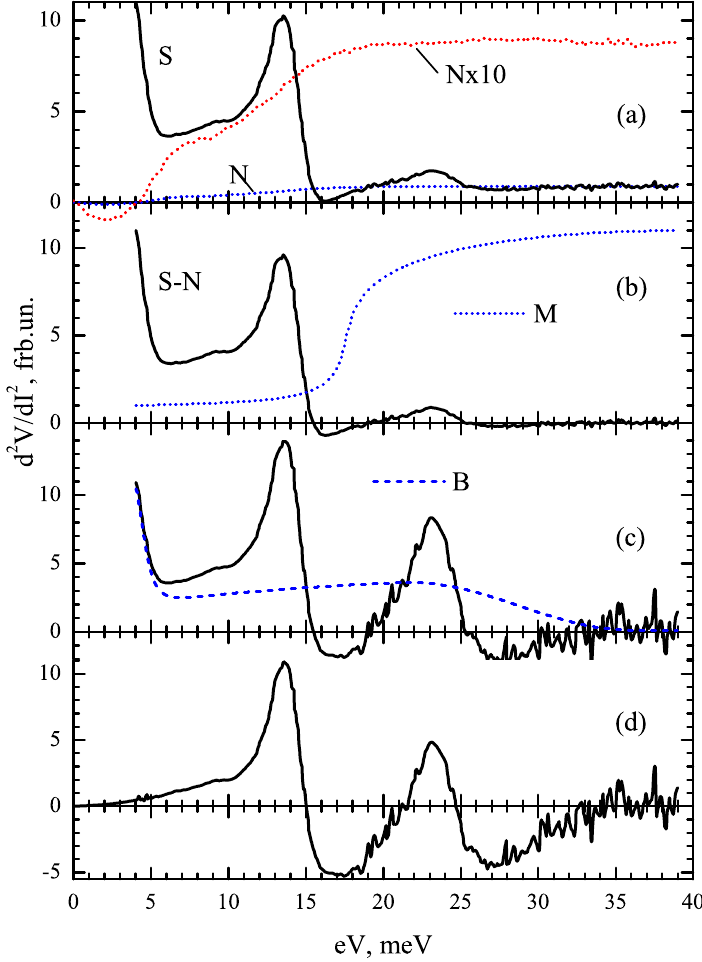}
\caption[]{\\(a) Second derivatives of I-V curves for an Nb point-contact, $R_{0}^{N}=17.5\ \Omega $, in the superconducting S and normal N states taken with the same
modulated voltage ($N\times 10$ is the scale multiplied by a factor of 10). \\(b) The
difference curve S-N; $M$ is the scaling curve analogous to the plot of Fig. \ref{Fig6}(c). \\(c) The spectrum after multiplication by the scaling curve; $B$ is the
background curve. \\(d) The spectrum after subtraction of the background curve.}
\label{Fig9}
\end{figure}
Here we show spectra of a "dirty" $Nb-Nb$ point-contact.
The contact resistance $R_{0}^{N}=17.5\ \Omega $. Figure \ref{Fig9}(a) shows the
second derivatives of the I-V curves in the normal and superconducting
states. It can be seen that the second derivative in
the normal state contains no peaks near the characteristic
phonon energies; only some small bends show up. This
behavior is typical of an amorphous material with disrupted
long-range order. It can be assumed that in the region of the
constriction the momentum mean free path is extremely
short. Taking the ratio ${{{\rho }_{300K}}}/{{{\rho }_{res}}}\;\sim 2$ at the center of the
contact and given that $\rho l=3,1\div 3,75\cdot {{10}^{-12}}\ \Omega \cdot c{{m}^{2}}$ (Refs.
\cite{Alekseevskii} and \cite{French}) and $\rho _{300K}^{Nb}=\text{14}.\text{2 1}{{0}^{-\text{6}}}\Omega \cdot cm$ \cite{Alekseevskii}, for $10\ K$ we
obtain a momentum mean free path of ${{l}_{i}}\sim 4.4\div 5.3\ nm$,
which is determined by impurities. Here the contact diameter
is $d\sim 11\ nm$. The coherence length for niobium is ${{\xi }_{0}}=38\ nm$ \cite{Antonova}, given that the momentum mean free path zeta, which
serves as a coherence length, is $\zeta =4\ nm$ $\left( 1/\zeta \ =1/{{\xi }_{0}}\ +1/{{l}_{i}}\  \right)$.
Therefore, since the contact is "dirty," we find that here the
length $\zeta$ is less than the contact diameter.
\begin{figure}[]
\includegraphics[width=8.5cm,angle=0]{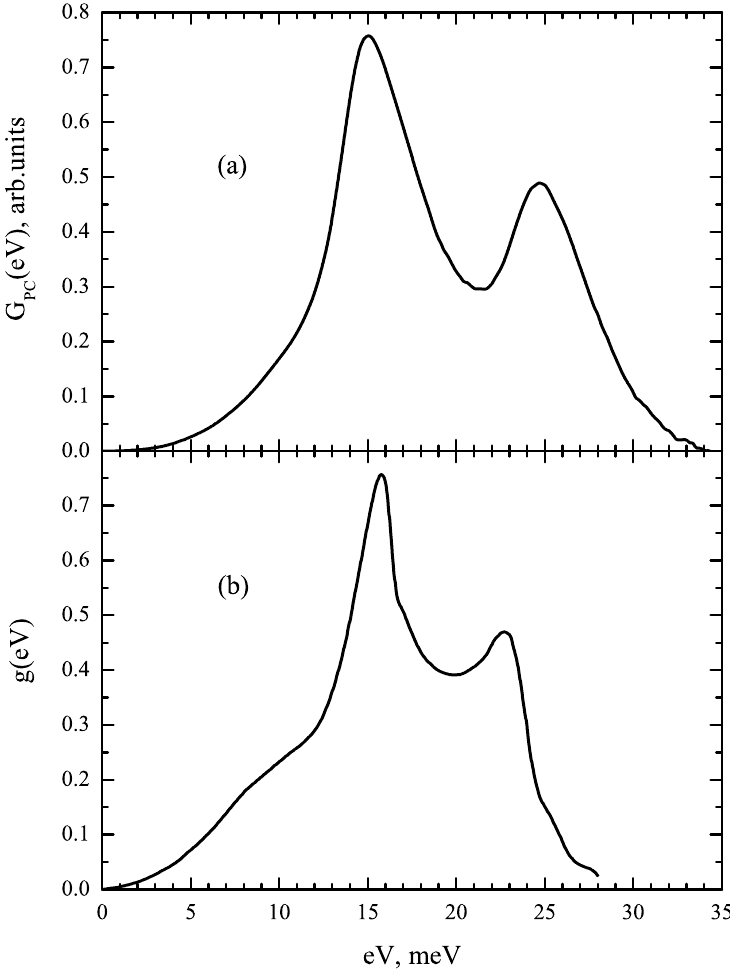}
\caption[]{\\(a) The EPI function for an $Nb$ point-contact obtained by integrating
the curve of Fig. \ref{Fig9}(d); \\(b) The EPI function for $Nb$ recovered from a tunnelling
spectrum \cite{Arnold}.}
\label{Fig10}
\end{figure}
Given the general form of the second derivative, we
may assume that suppression of the excess current is considerably
weaker in this case than it was for $NbSe_2$ in the previous
section. Since the excess current curve, which can be
used as a seed, is lacking in this case, the scaling curve is
drawn in by hand. Figure \ref{Fig10} shows the EPI function recovered
using the scaling curve compared with the EPI function
recovered from tunnelling data \cite{Arnold}.
\subsection{Discussion of results}
Experiments with, for example, tin (Figs. 2, 3, 4 of Ref. \cite{Bobrov1})
show that in the ballistic regime the superconducting contribution
to the spectrum is small compared to the spectrum in the
normal state and is no more than 20\%. As the elastic electron
mean free path falls off, the intensity of the point-contact
spectrum in the normal state decreases and becomes much
lower in the diffusive limit than in the ballistic limit. At the
same time, in the superconducting state, during the transition
from the ballistic to the diffusive limit the excess current falls
off by slightly less than a factor of two. Since the superconducting
contribution to the spectrum is proportional to the
magnitude of the excess current, we can expect that a reduction
in the elastic scattering mean free path will be accompanied
by a rise in the relative intensity of the superconducting
contribution compared to the spectrum in the normal state.
The above example of the characteristics of a "dirty" niobium
contact (Fig. \ref{Fig8}(a)) shows that the intensity of the spectrum for
the superconducting state in the phonon frequency range is an
order of magnitude higher than for the normal state. We note
the fundamental difference in the shape of the spectrum in the
normal state and the EPI function recovered from the
superconducting contribution to the spectrum; this confirms
the assertion in the Introduction regarding the spatial mismatch
of the volumes responsible for the formation of the
spectra in the normal and superconducting states. Since the
spectrum in the normal state is only produced by electrons
that have undergone backscattering, i.e., electrons that have
returned to the same electrode from which they originated
(see the Introduction), in "dirty" contacts the concentration
gradient of elastic scatterers which exists in all mechanically
formed contacts plays a very important role. The maximum
concentration of scatterers occurs at the boundary between the
electrodes and falls off inside the cantilevers. This means that
the diffusion of electrons after they are scattered on nonequilibrium
phonons takes place preferentially in the direction of
decreasing scatterer concentration. Because of this, the major
contribution to the spectrum in the normal state is from scattering
events immediately adjacent to the boundary between
the electrodes.

At the same time, any scattering process for nonequilibrium
phonons on Andreev electrons is efficient, since it
leads to a drop in the excess current. Thus, the spectrum in
the normal state will the formed directly at the boundary, in
the region where the scatterer concentration is highest, while
the superconducting contribution to the spectrum is formed
in a larger, cleaner region of the superconductor.
To summarize, we have the following:
\begin{enumerate}
\item {If the PC spectrum in the normal state and the EPI function
recovered from the superconducting contribution to
the current are fundamentally different in shape, then
they originate in geometrically different volumes of the
point-contact.}
\item {Since the conversion length for Andreev electrons into
Cooper pairs is the coherence length, for point-contacts
with ${d}\ge {{\xi }_{0}}$ the superconducting contribution to the
spectrum is formed in a volume smaller than the contact
diameter.}
\end{enumerate}

When items 1 and 2 are satisfied simultaneously, we are
dealing with a nonuniform point-contact in which impurities
and lattice distortions are concentrated near the boundary
between the electrodes. The spectrum in the normal state is
formed right at this boundary, while the spectrum in the
superconducting state develops in a volume extending from
the boundary by the coherence length, within which the crystal
lattice is not distorted so much.

When ${d}\ge {{\xi }_{0}}$, recovery of the EPI function from the
superconducting contribution requires that the reduction in
the relative intensity of the high-energy part of the phonon
spectrum be taken into account and a correction for the
decrease in the excess current must be included.
\section{Conclusions}
\begin{enumerate}
\item {It has been found that for point-contacts with diameters
larger than or on the order of the coherence length, the
relative magnitude of the negative phonon contribution
to the excess current is considerably greater than that for
ballistic contacts. This leads to substantial suppression
of the high-frequency peaks in the spectra for the superconducting
state. In order to recover the EPI function
from these spectra it is necessary to correct their intensities at high energies. For "dirty" $NbSe_2$ and $Nb$
point-contacts with ${d}\ge {{\xi }_{0}}$, which have no EPI spectrum
in the normal state, the EPI functions have been recovered
from the superconducting spectra.}
\item {The effect of a magnetic field on the current-voltage characteristics
and their derivatives for $S-c-S$ point-contacts of
$ErN{{i}_{2}}{{B}_{2}}C$ with ${d}\ge {{\xi }_{0}}$ has been studied. It was found that
in zero magnetic field and in fields close to those which
suppress superconductivity, when the size of the superconducting
gap can be neglected, the position of the peaks
in the differential resistance in the first derivatives of the
I-V curve coincides with the peaks of the Yanson EPI
spectra. In low fields the peaks are shifted toward lower
energies and in intermediate fields the peaks are split.}
\end{enumerate}
This work was supported by the National Academy of
Sciences of Ukraine as part of project FTs 3-19. The author
thanks A. V. Khotkevich for valuable consultations and
comments.


\begin{thebibliography}{}
\bibitem{Kulik} I. O. Kulik, A. N. Omel'yanchuk, and R. I. Shekhter, Fiz. Nizk. Temp. 3, 1543 (1977) [Sov. J. Low Temp. Phys. 3, 740 (1977)].

\bibitem{Khlus}  V. A. Khlus and A. N. Omel'yanchuk, \href{http://fntr.ilt.kharkov.ua/fnt/pdf/9/9-4/f09-0373r.pdf}{Fiz. Nizk. Temp.} 9, 373 (1983) [Sov. J. Low Temp. Phys. 9, 189 (1983)].

 \bibitem{Zaitsev} A. V. Zaitsev, \href{http://www.jetp.ac.ru/cgi-bin/dn/e_051_01_0111.pdf}{ZhETF} 78, 221 (1980).

\bibitem{Bobrov1} N. L. Bobrov, A. V. Khotkevich, G. V. Kamarchuk, and P. N. Chubov, \href{http://fntr.ilt.kharkov.ua/fnt/pdf/40/40-3/f40-0280r.pdf}{Fiz. Nizk. Temp.} 40, 280 (2014) [\href{http://dx.doi.org/10.1063/1.4869565}{Low Temp. Phys.} 40, 215 (2014)]; \href{http://arxiv.org/pdf/1405.6869.pdf}{arXiv:1405.6869}.

\bibitem{Bobrov2} N. L. Bobrov, V. V. Fisun, O. E. Kvitnitskaya, V. N. Chernobai, and I. K.
Yanson, \href{http://fntr.ilt.kharkov.ua/fnt/pdf/38/38-5/f38-0480r.pdf}{Fiz. Nizk. Temp.} 38, 480 (2012) [\href{http://dx.doi.org/10.1063/1.4709437}{Low Temp. Phys.} 38, 373 (2012)];  \href{http://arxiv.org/pdf/1207.6486.pdf}{arXiv:1207.6486}.

\bibitem{Bobrov3} N. L. Bobrov, V. N. Chernobay, Yu. G. Naidyuk, L. V. Tyutrina, I. K. Yanson, D. G. Naugle, and K. D. D. Rathnayaka, \href{http://fntr.ilt.kharkov.ua/fnt/pdf/36/36-10/f36-1228e.pdf}{Fiz. Nizk. Temp}. 36, 1228 (2010) [\href{http://dx.doi.org/10.1063/1.3521569}{Low Temp. Phys.} 36, 990 (2010)];  \href{http://arxiv.org/pdf/1006.5933.pdf}{arXiv:1006.5933}.

\bibitem{Bobrov4} N. L. Bobrov, V. N. Chernobay, Yu. G. Naidyuk, L. V. Tyutrina, D. G. Naugle, K. D. D. Rathnayaka, S. L. Bud'ko, P. C. Canfield, and I. K. Yanson, \href{http://iopscience.iop.org/article/10.1209/0295-5075/83/37003/pdf}{Europhys. Lett.} 83, 37003 (2008).

\bibitem{Naidyuk} Yu. G. Naidyuk and I. K. Yanson, \href{http://www.springer.com/gp/book/9780387212357}{Point-Contact Spectroscopy} (Springer,
New-York, 2005).

\bibitem{Shulga} S. V. Shulga, S.-L. Drechsler, G. Fuchs, K.-H. Mˆuller, K. Winzer, M.
Heinecke, and K. Krug, \href{http://dx.doi.org/10.1103/PhysRevLett.80.1730}{Phys. Rev. Lett.} 80, 1730 (1998).

\bibitem{Cho} B. K. Cho, P. C. Canfield, L. L. Miller, D. C. Johnston, W. P. Beyermann,
and A. Yatskar, \href{http://dx.doi.org/10.1103/PhysRevB.52.3684}{Phys. Rev. B} 52, 3684 (1995).

\bibitem{Skanthakumar} S. Skanthakumar and J. W. Lynn, \href{http://www.sciencedirect.com/science/article/pii/S0921452698007303}{Physica B} 259–261, 576 (1999).

\bibitem{Yanson1} I. K. Yanson, L. F. Rybal'chenko, V. V. Fisun, N. L. Bobrov, M. A.
Obolenskii, M. V. Kosmyna, and V. P. Seminozhenko, \href{http://fntr.ilt.kharkov.ua/fnt/pdf/14/14-11/f14-1157r.pdf}{Fiz. Nizk. Temp.}
14, 1157 (1988) [Sov. J. Low Temp. Phys. 14, 639 (1988)].

\bibitem{Bud'ko} S. L. Bud'ko and P. C. Canfield, \href{http://dx.doi.org/10.1103/PhysRevB.61.R14932}{Phys. Rev. B} 61, R14932 (2000).

\bibitem{Yanson2} I. K. Yanson, L. F. Rybal'chenko, N. L. Bobrov, and V. V. Fisun, \href{http://fntr.ilt.kharkov.ua/fnt/pdf/12/12-5/f12-0552r.pdf}{Fiz.
Nizk. Temp.} 12, 552 (1986) [Sov. J. Low Temp. Phys. 12, 313 (1986)]; \href{http://arxiv.org/pdf/1512.00684.pdf}{arXiv:1512.00684}.

\bibitem{Bobrov5} N. L. Bobrov, L. F. Rybal'chenko, M. A. Obolenskii, and V. V. Fisun, \href{http://fntr.ilt.kharkov.ua/fnt/pdf/11/11-9/f11-0925r.pdf}{Fiz.
Nizk. Temp.} 11, 925 (1985) [Sov. J. Low Temp. Phys. 11, 510 (1985)] .

\bibitem{Alekseevskii} N. E. Alekseevskii, V. I. Nizhanovskii, and K. H. Bertel, FMM 37, 63
(1974).

\bibitem{French} R. A. French, \href{http://www.sciencedirect.com/science/article/pii/S0011227568800074}{Cryogenics} 8, 301 (1968).

\bibitem{Antonova}I. Yu. Antonova, V. M. Zakosarenko, E. V. Il'ichev, V. I. Rozenflants, and
V. A. Tulin, \href{http://journals.ioffe.ru/jtf/1990/03/p135-140.pdf}{ZhTF} 60, 135 (1990).

\bibitem{Arnold} G. B. Arnold, J. Zasadzinski, J. W. Osmun, and E. L. Wolf, \href{http://link.springer.com/article/10.1007/BF00117117}{J. Low Temp.
Phys.} 40, 225 (1980).
\end{thebibliography}
\end{document}